\documentclass[3p,times]{article}

\usepackage{amsmath}
\usepackage{amsthm}
\usepackage{booktabs}

\usepackage{amssymb}

\usepackage[figuresright]{rotating}
\usepackage{tcolorbox}

\usepackage[font=footnotesize,labelfont=bf]{caption}
\usepackage[font=footnotesize,labelfont=bf]{subcaption}

\usepackage[english]{babel}
\title{On the Evolution of Covid-19 in Italy: \\a Follow up Note}

\author{Giuseppe Dattoli, Emanuele Di Palma,\\ Silvia Licciardi\footnote{Corresponding author: silviakant@gmail.com, silvia.licciardi@enea.it, orcid 0000-0003-4564-8866, tel. nr: +39 06-94005421.}, Elio Sabia  \\[1.3ex]
	ENEA - Frascati Research Center, Via Enrico Fermi 45, 00044, \\Frascati, Rome, Italy  
}

\date{\today}
\begin{document}

	\maketitle

	\begin{abstract}
	In a previous note we made an analysis of the spreading of the COVID disease in Italy. We used a model based on the logistic and Hubbert functions, the analysis we exploited has shown limited usefulness in terms of predictions and failed in fixing fundamental indications like the point of inflection of the disease growth. In this note we elaborate on the previous model, using multi-logistic models and attempt a more realistic analysis.
	\end{abstract}
	
	\section{Introduction}	

The Covid-19 pandemic disease is bringing elements of novelty baffling for politicians, $MD$'s and epidemic analyzers. \\

It has already been stressed that, in absence of any anti-viral strategy, the only defense towards the spreading of the illness is the Nation lockdown, a policy difficult to implement in Italy. It has undergone different phases and lack of effective decisions, while the infection was raging in Italy and attacking the rest of Europe.\\

The public health structures have suffered from an un-precedent stress in terms of people to care and of casualties. Regarding this last point, in Italy the percentage of deaths seems to be larger than abroad, but this might be result of absence of an accurate sampling of the positive cases.\\

The lack of informations on the infecting capabilities of the virus and other uncertainties associated with a clear understanding of how the infection developed during the early stages of its spreading and a poor knowledge on the real entity of the ``submerged" cases as well, made any attempt to fix the peak of the distribution of the infected/day denied by the facts themselves.\\

In a previous note, by the present group of Authors \cite{covid}, two paradigmatic tools have been exploited to study the evolution of the illness: \\

\begin{figure}[h]
 	\centering
 	\includegraphics[width=0.6\linewidth]{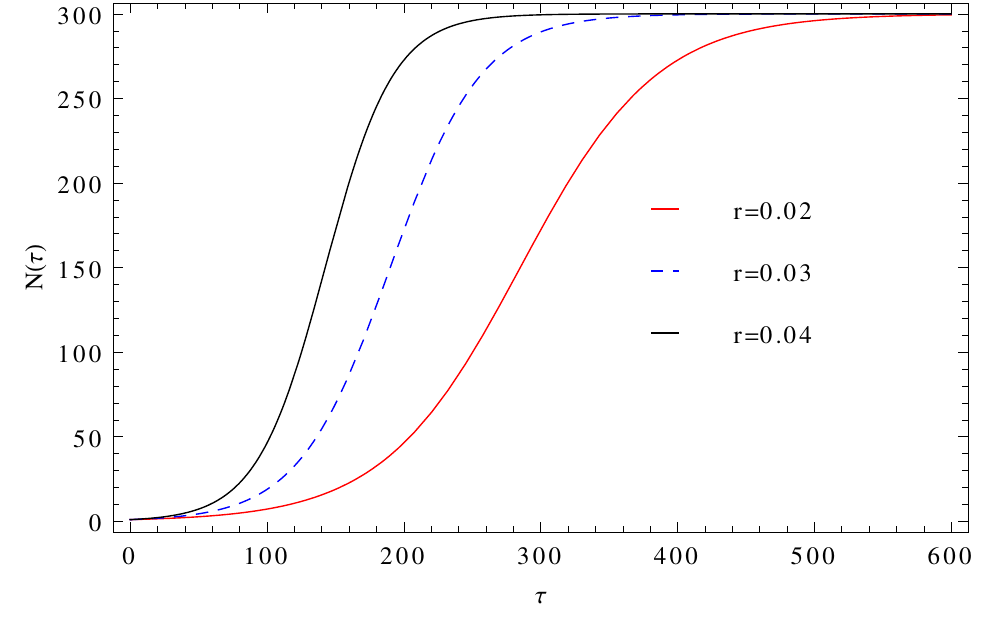}
 	\caption{Growth of infected individual vs. $\tau$ for $K=300$ and different values of the growth rate $r$.}
 	\label{fig:codfig1}
 \end{figure}

\noindent A) The logistic function \cite{Cramer,Weisstei,Fisher} ($LF$) (Fig. \ref{fig:codfig1}) describing the evolution of a given population $N(\tau)$ of $N_0$ individuals at $\tau\!=\!0$ (in the present case \textbf{infected people}) in an environment with carrying capacity $K$ and growth rate $r$, is specified by 
\begin{equation}\label{nt}
N(\tau)=N_0\dfrac{e^{r\tau}}{1+\dfrac{N_0}{K}(e^{r\tau}-1)},
\end{equation}
 where $\tau$ is the time, measured in some units to be specified. \\
 
\begin{figure}[h]
	\centering
	\includegraphics[width=0.5\linewidth]{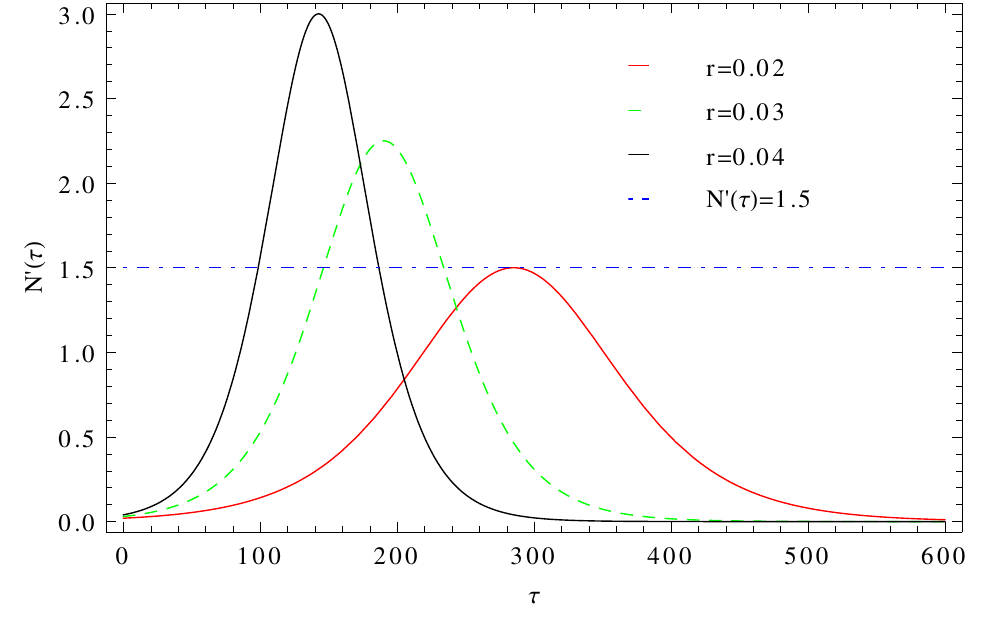}
	\caption{Hubbert curve vs. $\tau$ for $K=300$ and different values of the growth rate $r$ and a hypothetical threshold rate.}
	\label{fig:codfig2}
\end{figure}

\noindent B) The Hubbert curve \cite{Deffeyes} ($HC$), namely the derivative of the $LF$, yielding the number of infections per unit time, i.e. 
\begin{equation}\label{key}
N'(\tau)=
\frac{e^{rt} rN_0(K-N_0) }{K\left( 1+\dfrac{N_0}{K}(e^{r t}-1)\right)^2 }.
\end{equation}
 It is a bell shaped curve (Fig. \ref{fig:codfig2}) with the maximum located at
\begin{equation}\label{key}
\tau^ *=\ln\left(\sqrt[r]{\dfrac{K}{N_0} -1}\right) .
\end{equation}
In correspondence of which the infected rate is
 \begin{equation}\label{key}
 N'(\tau^ *)=\dfrac{r K}{4}
 \end{equation}
corresponding to a total number of infected
\begin{equation}\label{key}
N(\tau^*)=\dfrac{K}{2}.
\end{equation}

\begin{figure}[h]
	\begin{subfigure}{0.49\textwidth}
		\includegraphics[width=.9\linewidth]{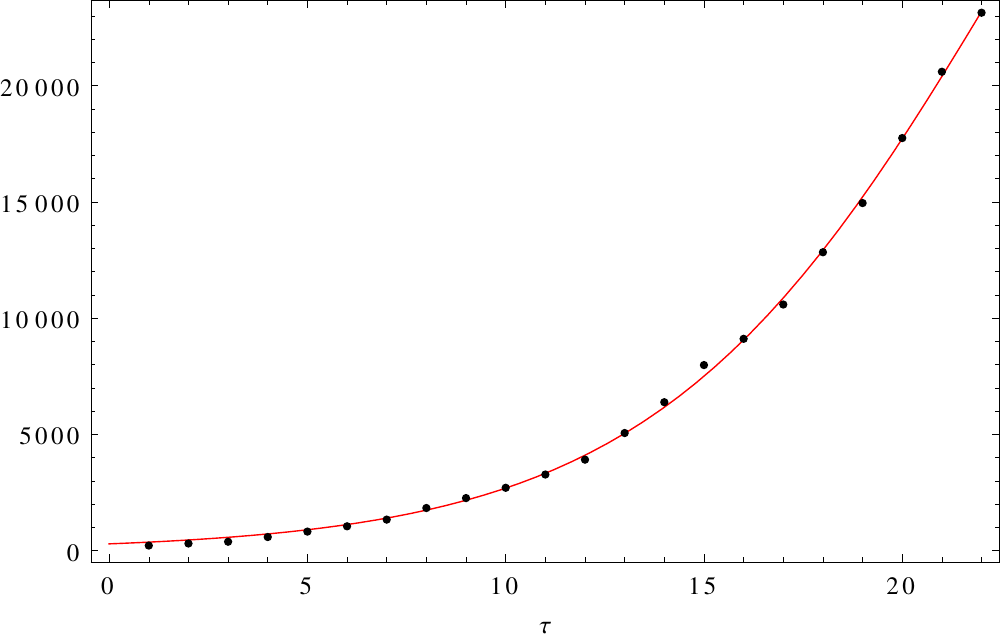}
		\caption{Real data of infected Italian individuals (black dotted) and Logistic equation \eqref{nt}.}
		\label{fig:codfig3a}
	\end{subfigure}\;
	\begin{subfigure}{0.49\textwidth}
		\includegraphics[width=.9\linewidth]{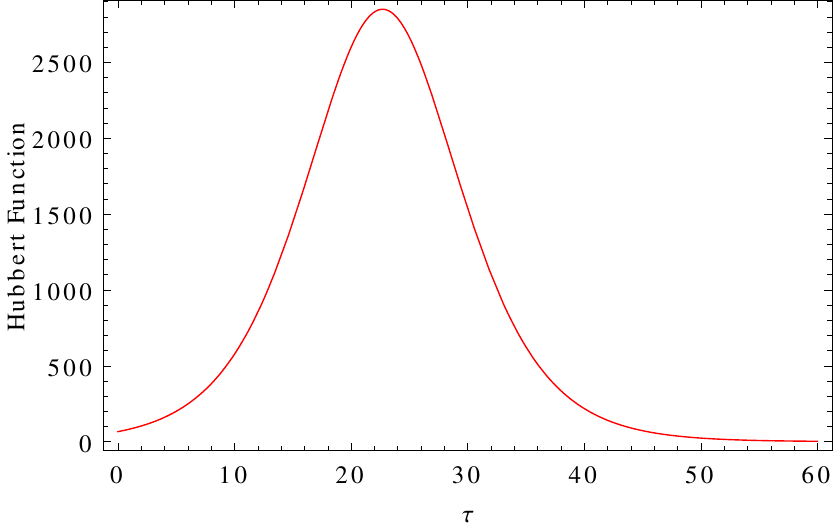}
		\caption{Hubbert function vs. $\tau$}
		\label{fig:codfig3b}
	\end{subfigure}
	\caption{Fitted functions in the period February 24 -- March 16, 2020,  for $N_0=294$, $r=0.2264$, $K=50346$.}
	\label{fig:codfig3}
\end{figure}

\begin{figure}[h]
	\begin{subfigure}{0.49\textwidth}
		\includegraphics[width=1.\linewidth]{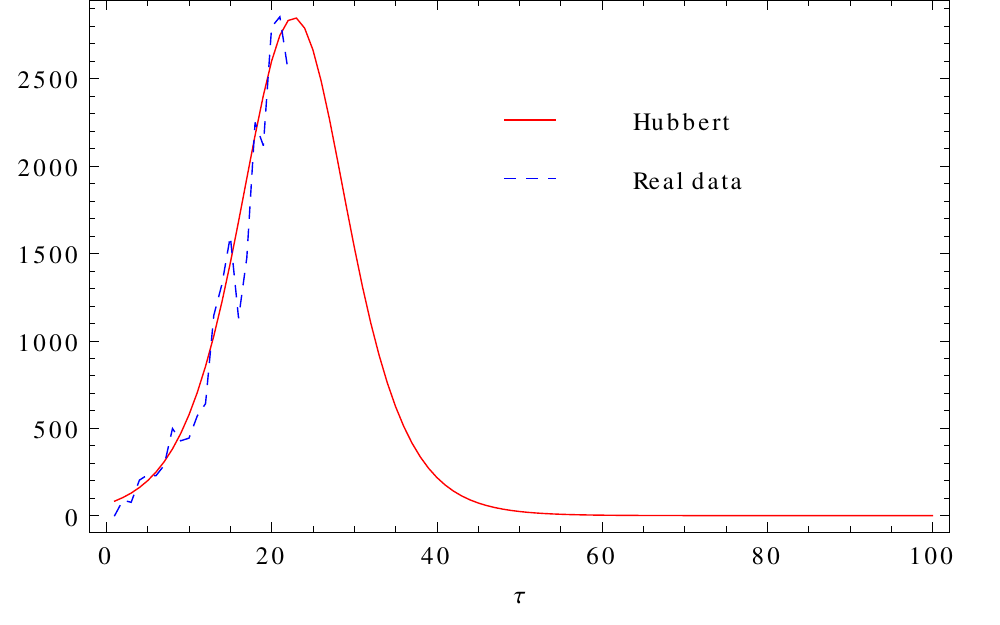}
		\caption{February 24-March 16.}
		\label{fig:codfig4a}
	\end{subfigure}\;
	\begin{subfigure}{0.49\textwidth}
		\includegraphics[width=1.\linewidth]{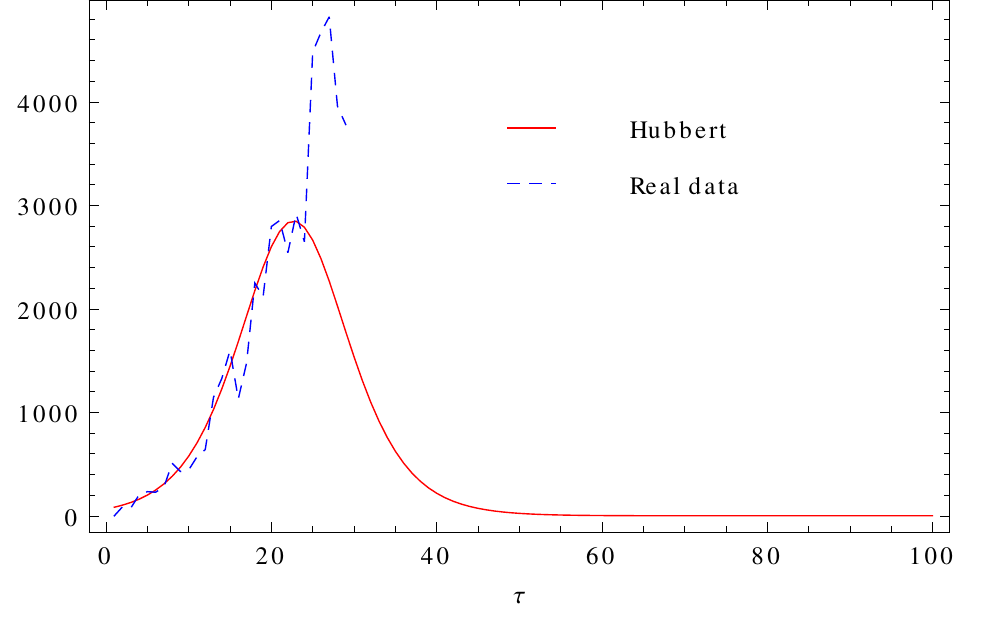}
		\caption{February 24-March 23.}
		\label{fig:codfig4b}
	\end{subfigure}
	\caption{Comparison between Hubbert curve, representing the number of Covid-19 positive per day, obtained from the fitted equations, and daily increment from the registered data.}
	\label{fig:codfig4}
\end{figure}

The analysis of the data provided by the Italian Ministry of health before March 19  where compatible with the scenario summarized in Figs. \ref{fig:codfig3} and \ref{fig:codfig4a}, namely the saturation of the infection by the end of April, the peak of infection rate around 17 of March.\\

The officially presented data on March 20 upset this ``reassuring" scenario and modified Fig. \ref{fig:codfig4a} as reported in Fig. \ref{fig:codfig4b}. The latter being characterized by an apparently anomalous behavior, dominated by an increase which mocks any every forecast based on a “simple” logistic model.\\
 
\begin{figure}[h]
	\centering
	\includegraphics[width=0.5\linewidth]{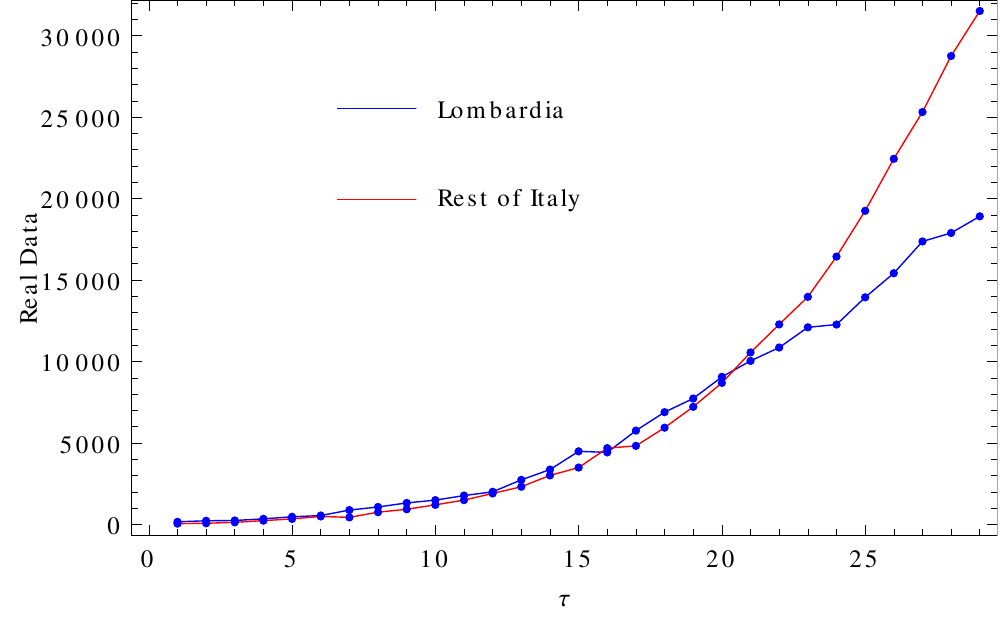}
	\caption{Number of infected since February 24.}
	\label{fig:codfig5}
\end{figure}

As already stressed in Ref. \cite{covid}, the analysis of the data at national territory level had been developed with the bias that the barycenter of illness was shifted towards the north of Italy. What was going to happen in those days has been that the cases from the rest of Italy were surpassing those in Lombardia (see Fig. \ref{fig:codfig5}). This imposes a new scenario in terms of statistical analysis as discussed in the forthcoming sections.
 
\section{Covid Bi-Logistic Models}

In Ref. \cite{covid} we underscored the possibility that the logistic model might be not suitable for a description at national level in view of various in-homogeneities of the distribution of the infection and for the delay in the propagation, presumably also mediated by the massive transfer of people from north to south of Italy. \\

\begin{figure}[h]
	\centering
	\includegraphics[width=0.5\linewidth]{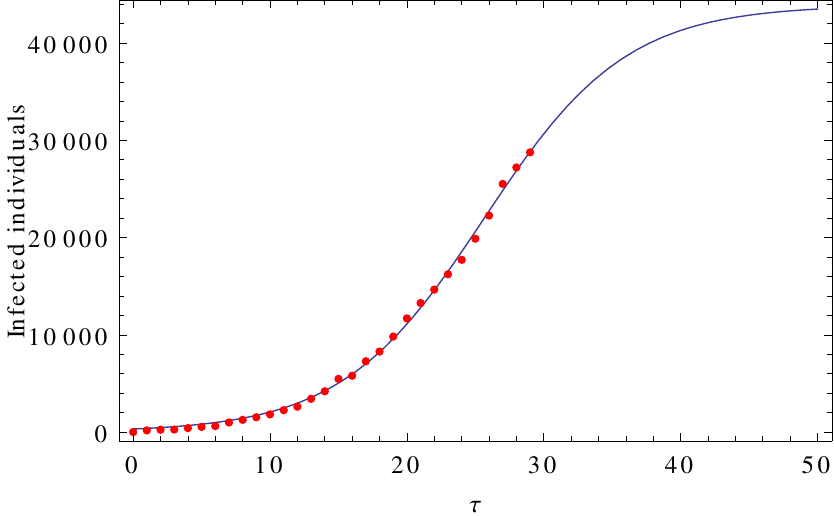}
	\caption{``Regione Lombardia" (red dot) and logistic interpolation $N(\tau)$ (continuous blue line) (February 24-March 24).}
	\label{fig:codfig6}
\end{figure}

\noindent Before considering a more elaborated point of view, we consider the data from ``Regione Lombardia" only, where we have reported the relevant logistic curve (Fig. \ref{fig:codfig6}). It should be noted that the curves are relevant to sum of casualties and infected. \\
  
\begin{figure}[h]
	\centering
	\includegraphics[width=0.5\linewidth]{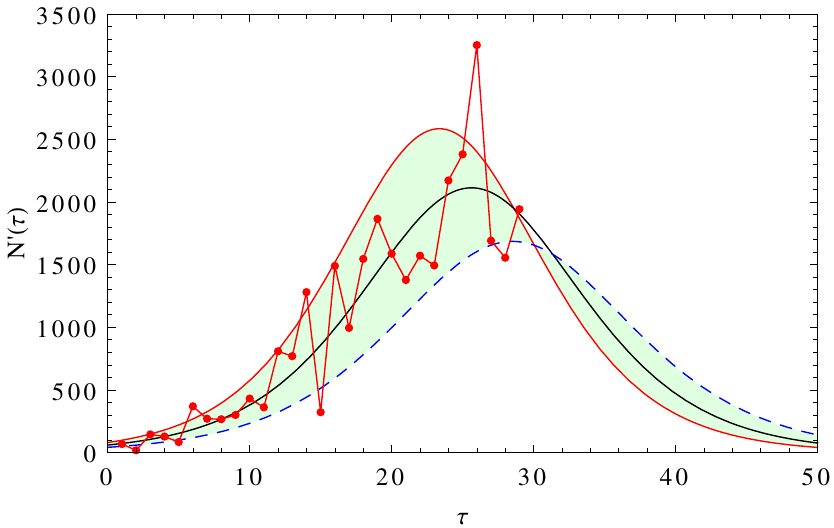}
		\caption{``Regione Lombardia" data and Hubbert function with region of $95\%$ of confidence (February 24-March 24).}
	\label{fig:codfig7}
\end{figure}

The fit of the Hubbert with $95\%$ of confidence band is given in Fig. \ref{fig:codfig7}, which displays three possible scenarios for the behavior of the infected and deceased/day. According to the previous forecasting, the peak should be reached in the next days. The lower curve predicts a peak by the end of march. \\

\begin{figure}[h]
	\begin{subfigure}{0.49\textwidth}
		\centering
		\includegraphics[width=.9\linewidth]{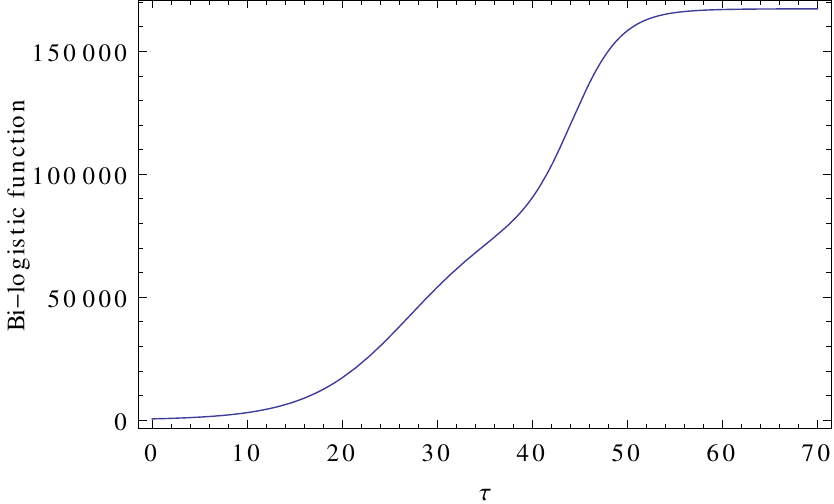}
		\caption{Bi-logistic evolution Eq. \eqref{nt}.}
		\label{fig:codfig8a}
	\end{subfigure}\;
	\begin{subfigure}{0.49\textwidth}
		\centering
		\includegraphics[width=.9\linewidth]{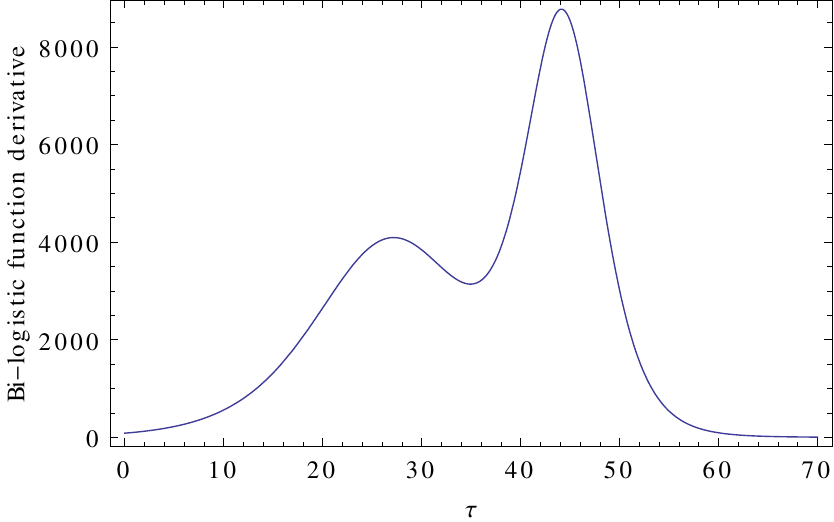}
		\caption{Bi-logistic diffusion in the time.}
		\label{fig:codfig8b}
	\end{subfigure}\;
	\caption{Bi-logistic evolution and diffusion in lag time of $31$ days.}
	\label{fig:codfig8}
\end{figure}

In order to extend the analysis to the national territory, we have elaborated a different strategy using the \textit{bi-logistic analysis}. We have therefore considered the incoherent sum of two logistics \cite{Meyer,MeyerJ}. They are characterized by different growth rates and carrying capacities. The time differences $\tau_2-\tau_1$ represents the time lag between the starting point of the two evolutions 
\begin{equation}\label{nt}
N(\tau)=N_{0,1}\dfrac{e^{r_1(\tau-\tau_1)}}{1+\dfrac{N_{0,1}}{K_1}(e^{r_1(\tau-\tau_1)}-1)}+
N_{0,2}\dfrac{e^{r_2(\tau-\tau_2)}}{1+\dfrac{N_{0,2}}{K_2}(e^{r_2(\tau-\tau_2)}-1)}.
\end{equation}
In Fig. \ref{fig:codfig8} we have reported the results of a $6$ parameters fit $(N_{0,i}, r_i, \tau_i)$, $i=1,2$, and assuming a lag time of $31$ days, corresponding to the difference in time between the (official) start of the disease and the crossing time between the two curves in Fig. \ref{fig:codfig5}.\\

The fit displays almost equivalent $N_0$ and $K$, but different growth rates. Regarding the associated Hubbert curve we obtain a plot exhibiting two peaks, with a delay between the two. This is a possible scenario, albeit questionable since it assumes that the rest of Italy starts to contribute to the counting after a significant time lag.\\

\begin{figure}[h]
	\centering
	\includegraphics[width=0.5\linewidth]{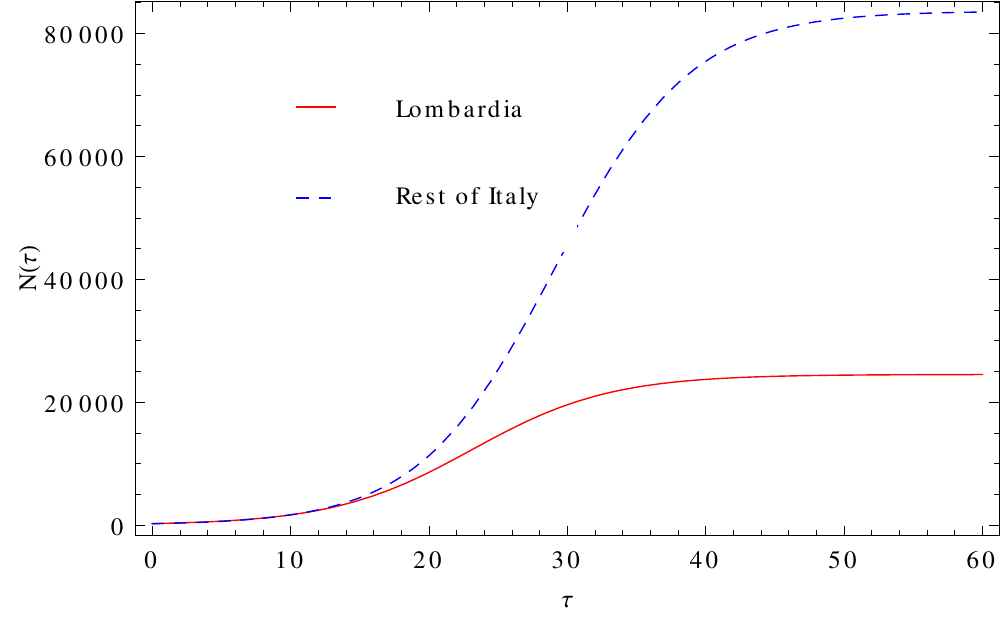}
	\caption{Fitted Logistic equations according to real data (February 24-March 23).}
	\label{fig:codfig9}
\end{figure}

\begin{figure}[h]
	\centering
	\includegraphics[width=0.6\linewidth]{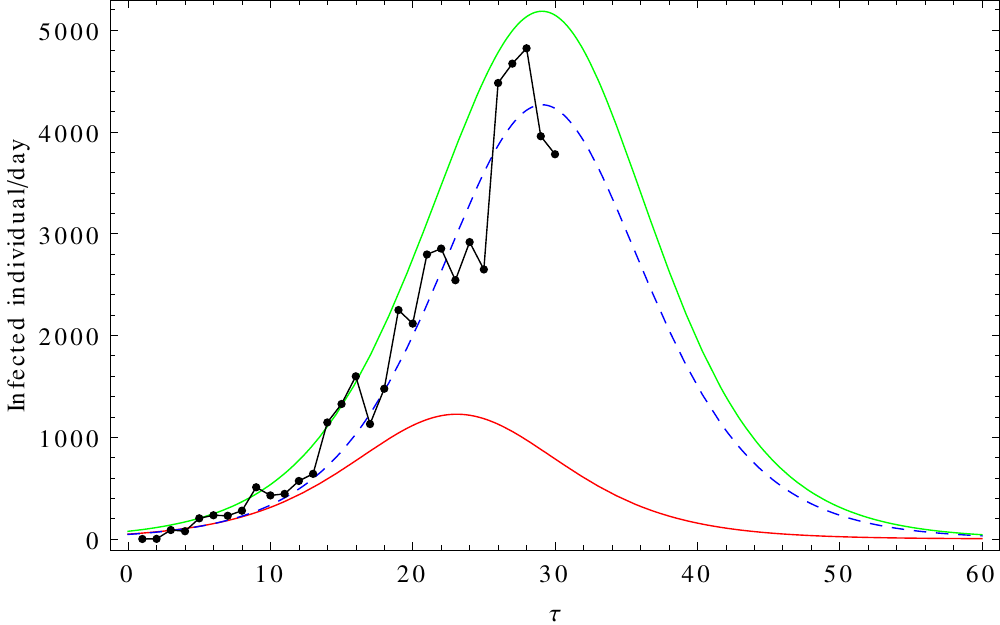}
	\caption{Hubbert curves: Regione Lombardia (red curve), Rest of Italy (blue dot curve), Bi-logistic of the two curves (green), National official data (black broken line).}
	\label{fig:codfig10}
\end{figure}

An alternative strategy is that summarized in Figs. \ref{fig:codfig9} and \ref{fig:codfig10}, in which we have treated Lombardia and the rest of Italy as separated entities. The fit has been done using two independent logistics which have been summed ``incoherently", thus getting two distinct Hubbert curves and the relevant sum, exhibiting the peak in the next few days.\\


\section{Final Comments}

\indent In this follow up we have exploited a larger number of data on Covid spreading and evolution in Italy, to gain a more accurate scenario on the present status and how it may evolve.\\
\indent Regarding this last point, many caveats are in order mainly with the understanding of the consistency of the submerged positives and how they may evolve in the next days. At moment it is not possible to have a clear understanding of the impact of the restrictions on the evolution of the illness diffusion. The data from the single Italian regions may be, within this respect, instructive.\\

\begin{figure}[h]
	\begin{subfigure}{0.49\textwidth}
		\centering
		\includegraphics[width=.9\linewidth]{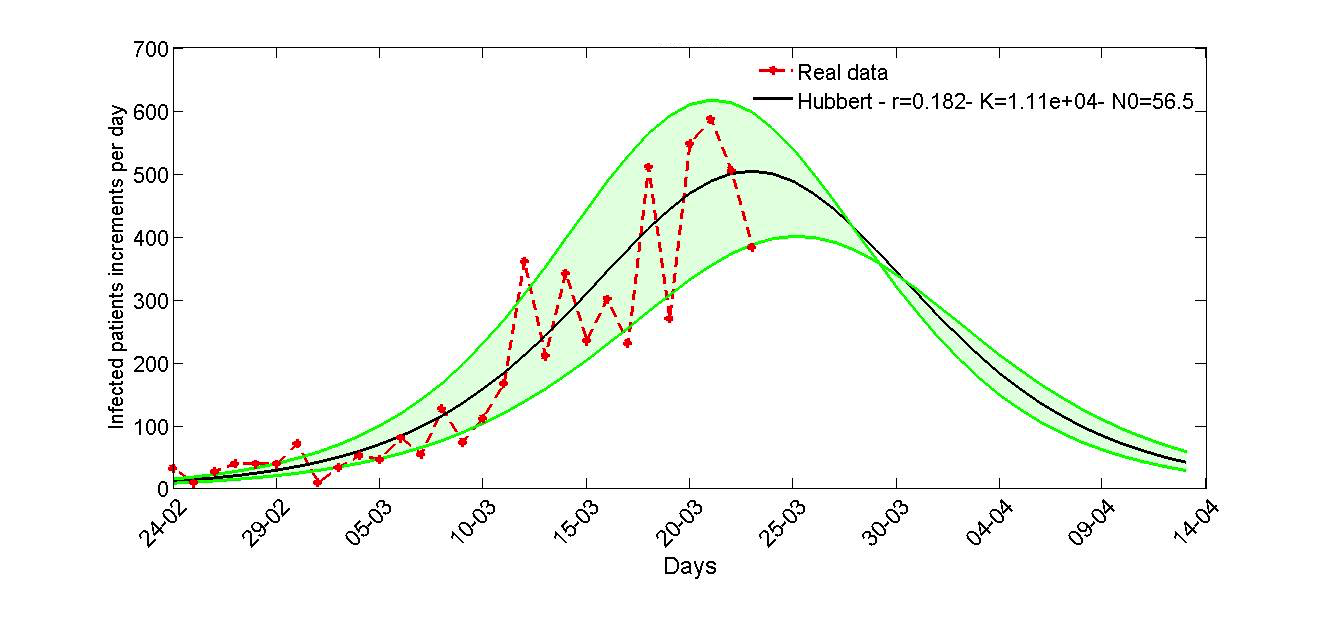}
		\caption{``Regione Veneto".}
		\label{fig:codfig11a}
	\end{subfigure}\;
	\begin{subfigure}{0.49\textwidth}
		\centering
		\includegraphics[width=.8\linewidth]{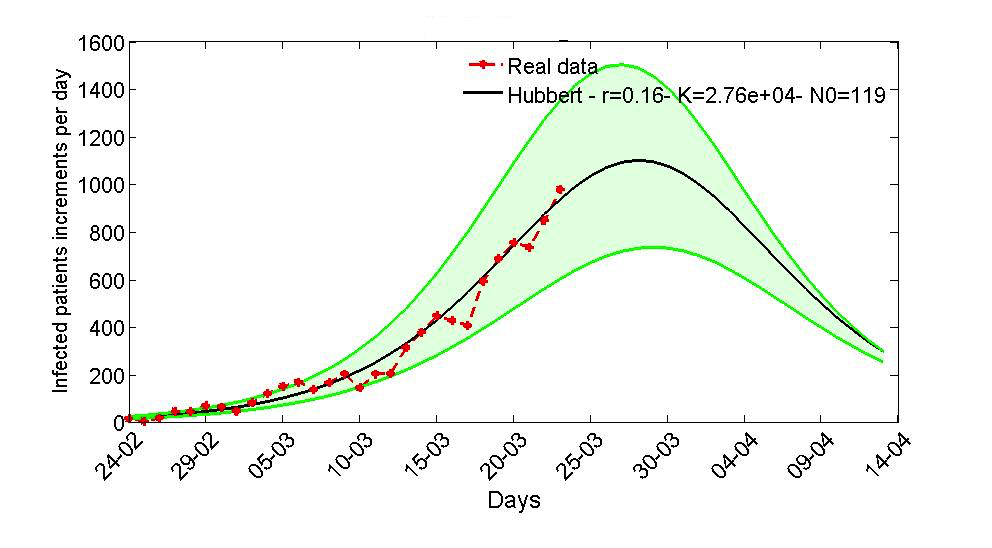}
		\caption{``Regione Emilia Romagna".}
		\label{fig:codfig11b}
	\end{subfigure}
	\caption{Selected sample of regions data and Hubbert function with region of $95\%$ of confidence.}
	\label{fig:codfig11}
\end{figure}

\begin{figure}[h]
\begin{subfigure}{0.48\textwidth}
		\centering
		\includegraphics[width=.8\linewidth]{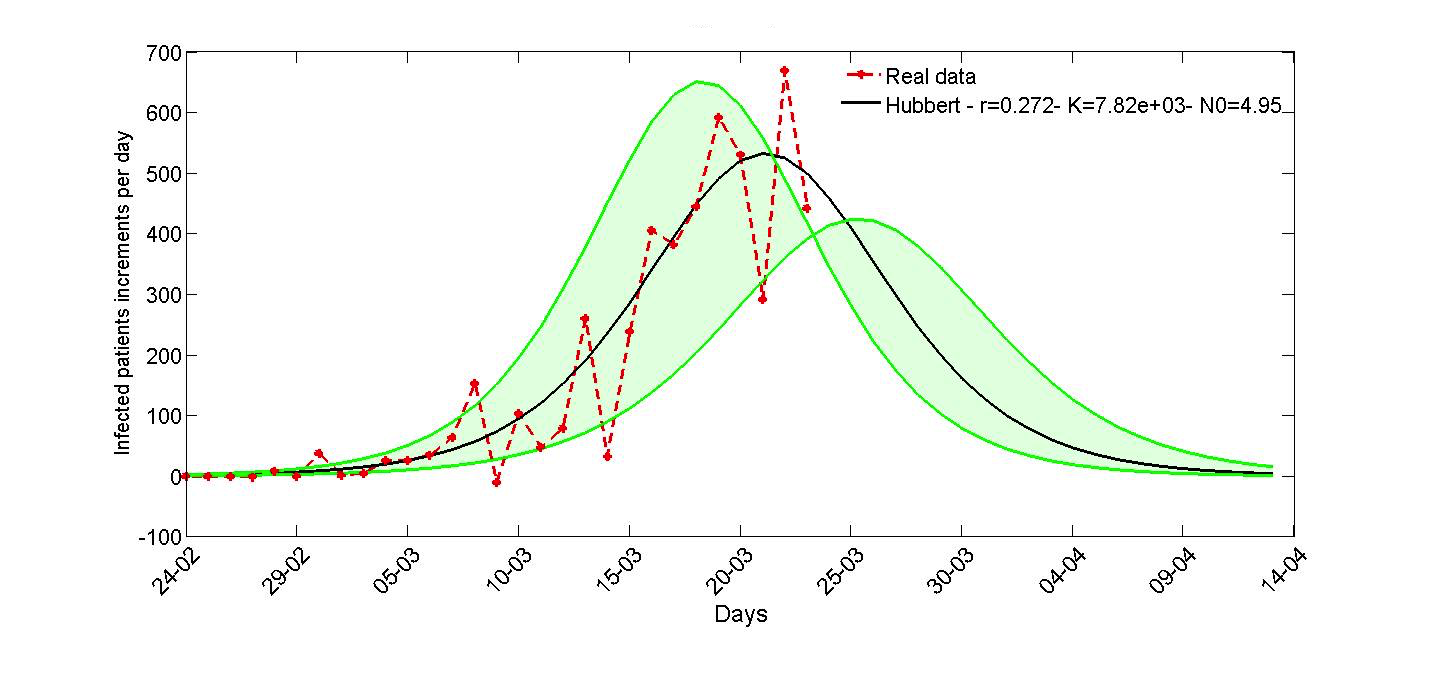}
		\caption{``Regione Piemonte".}
		\label{fig:codfig11c}
	\end{subfigure}
\begin{subfigure}{0.48\textwidth}
		\centering
		\includegraphics[width=.8\linewidth]{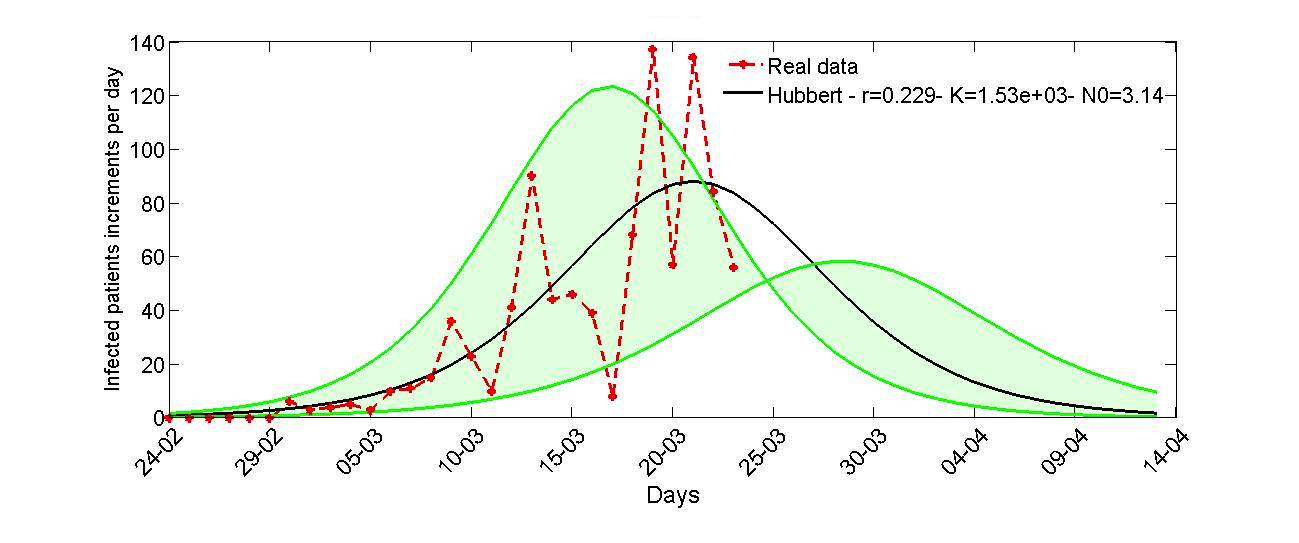}
		\caption{``Regione Friuli".}
		\label{fig:codfig11d}
	\end{subfigure}
\begin{subfigure}{0.48\textwidth}
		\centering
		\includegraphics[width=.8\linewidth]{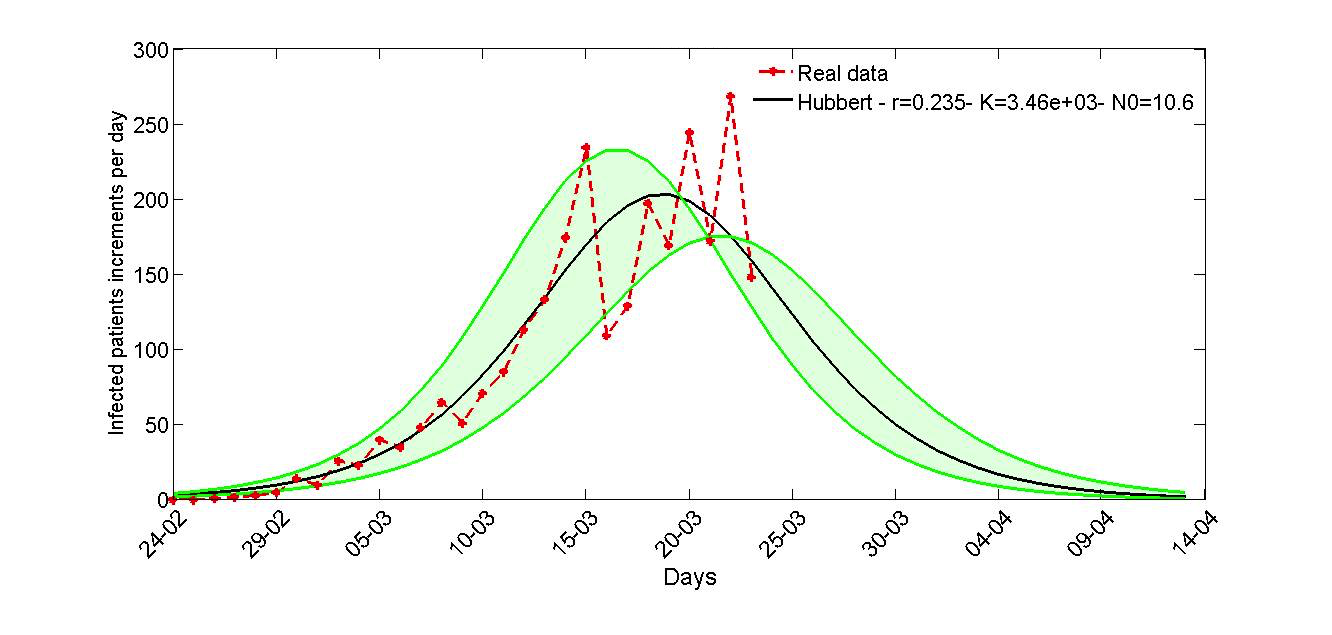}
	\caption{``Regione Marche".}
	\label{fig:codfig11e}
   \end{subfigure}
\begin{subfigure}{0.48\textwidth}
	\centering
	\includegraphics[width=.8\linewidth]{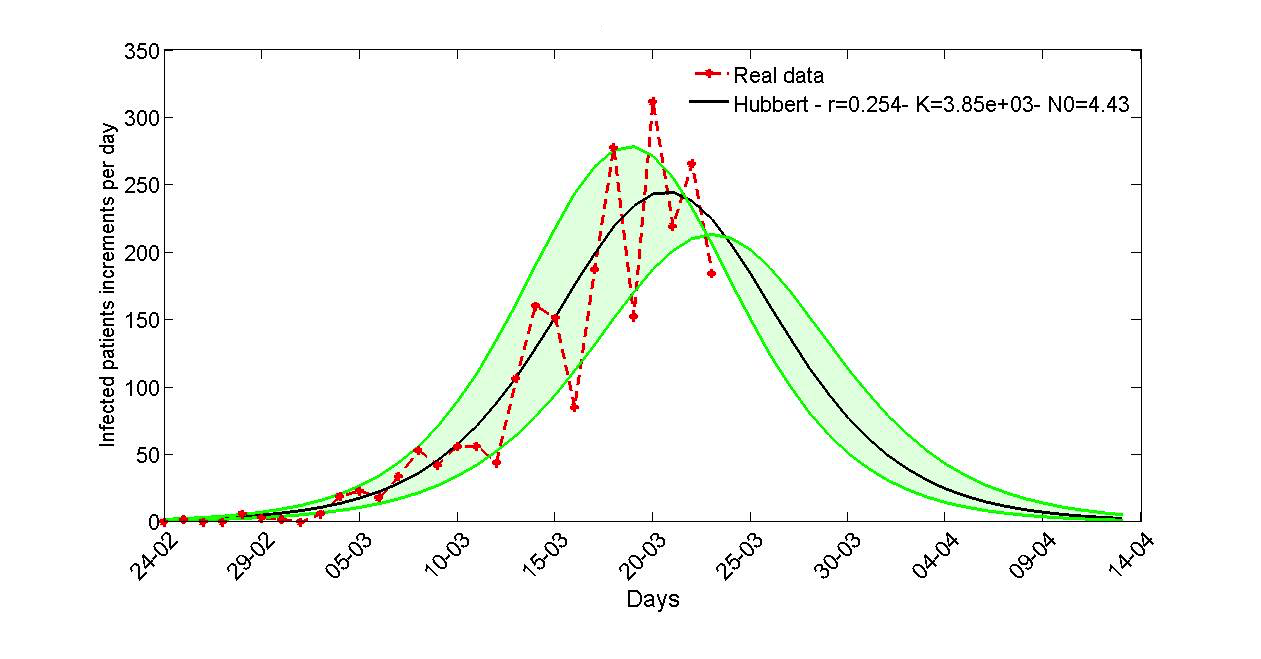}
	\caption{``Regione Toscana".}
	\label{fig:codfig11f}
   \end{subfigure}
	\begin{subfigure}{0.48\textwidth}
	\centering
	\includegraphics[width=.8\linewidth]{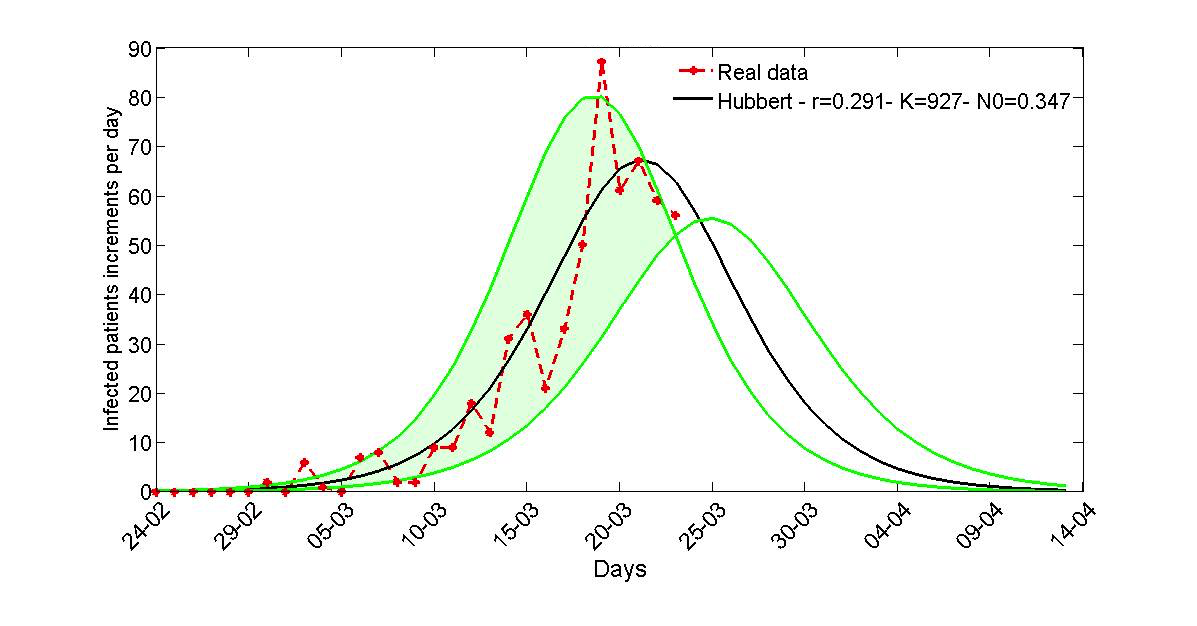}
	\caption{``Regione Umbria".}
	\label{fig:codfig11g}
\end{subfigure}
\begin{subfigure}{0.48\textwidth}
	\centering
	\includegraphics[width=.8\linewidth]{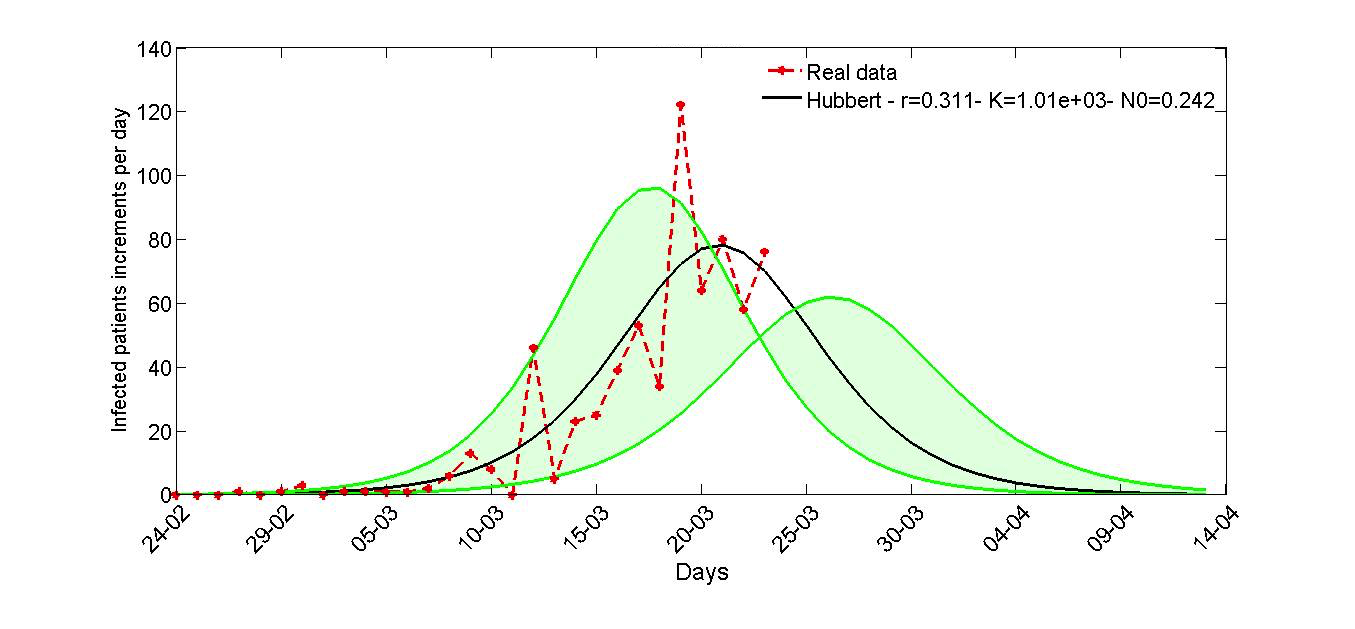}
	\caption{``Regione Abruzzo".}
	\label{fig:codfig11h}
\end{subfigure}
\begin{subfigure}{0.48\textwidth}
	\centering
	\includegraphics[width=.8\linewidth]{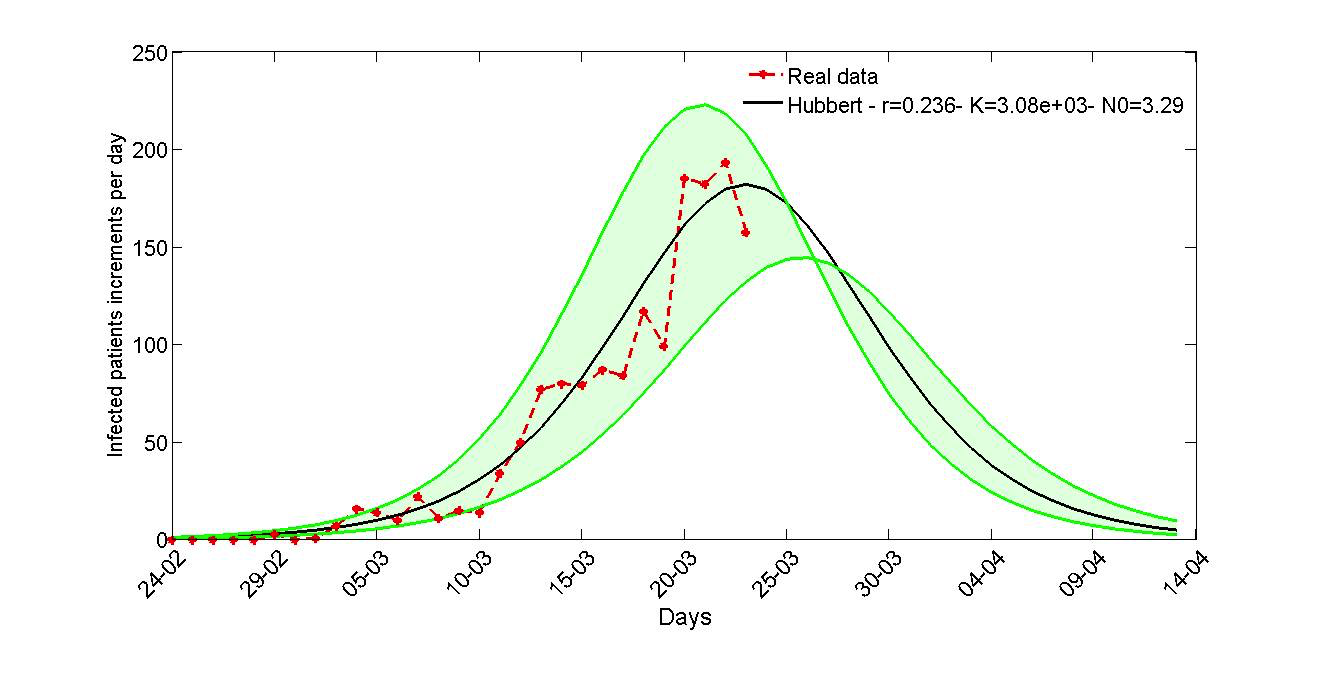}
	\caption{``Regione Lazio".}
	\label{fig:codfig11i}
\end{subfigure}\;\;
\begin{subfigure}{0.48\textwidth}
	\centering
	\includegraphics[width=.8\linewidth]{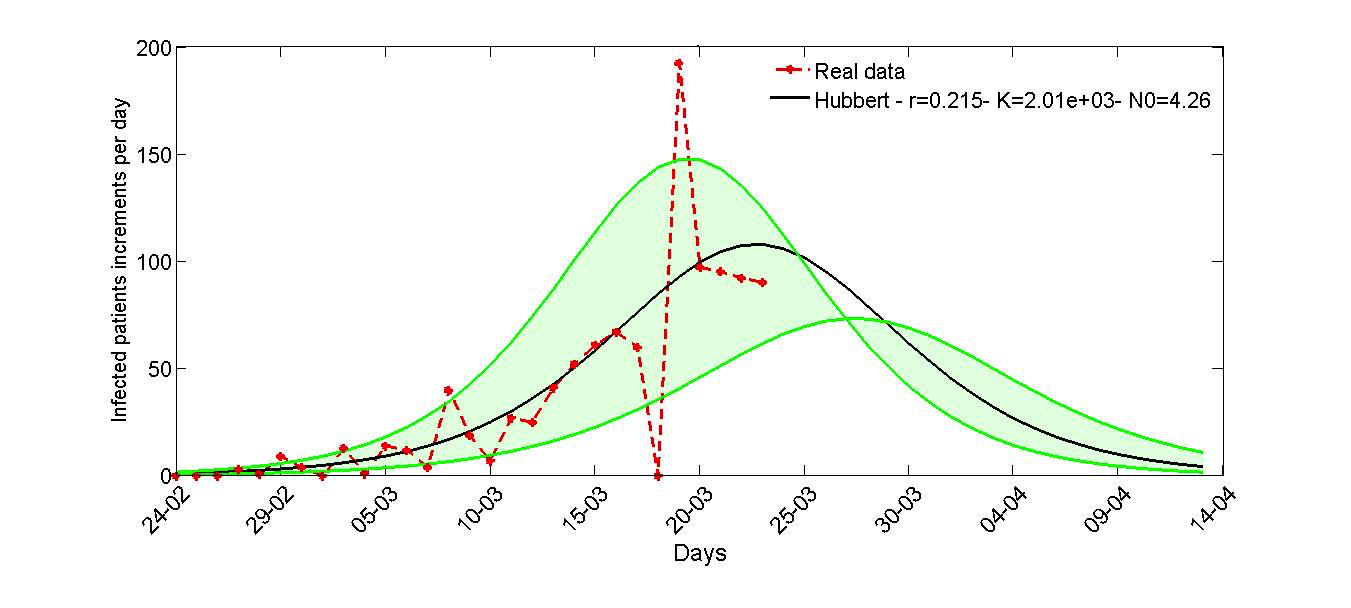}
	\caption{``Regione Campania".}
	\label{fig:codfig11l}
\end{subfigure}
	\caption{Selected sample of regions data and Hubbert function with region of $95\%$ of confidence.}
\label{fig:codfig11bis}
\end{figure}

\begin{figure}[h]
\begin{subfigure}{0.49\textwidth}
		\centering
		\includegraphics[width=.9\linewidth]{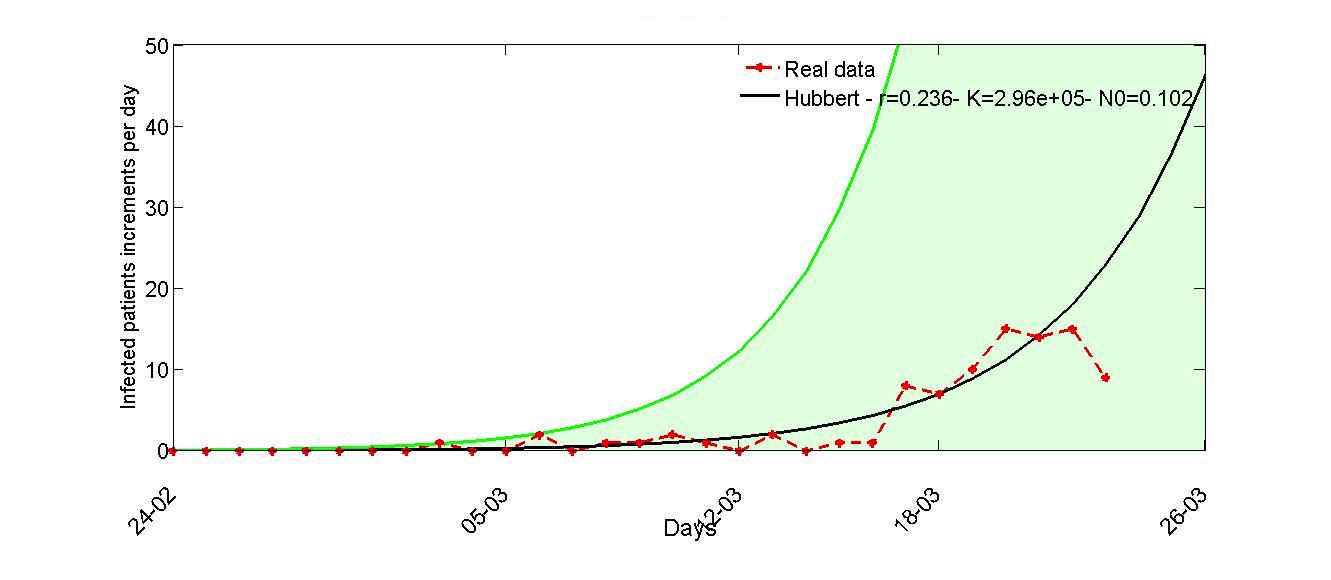}
		\caption{``Regione Basilicata".}
		\label{fig:codfig11m}
	\end{subfigure}\;
\begin{subfigure}{0.49\textwidth}
		\centering
		\includegraphics[width=.9\linewidth]{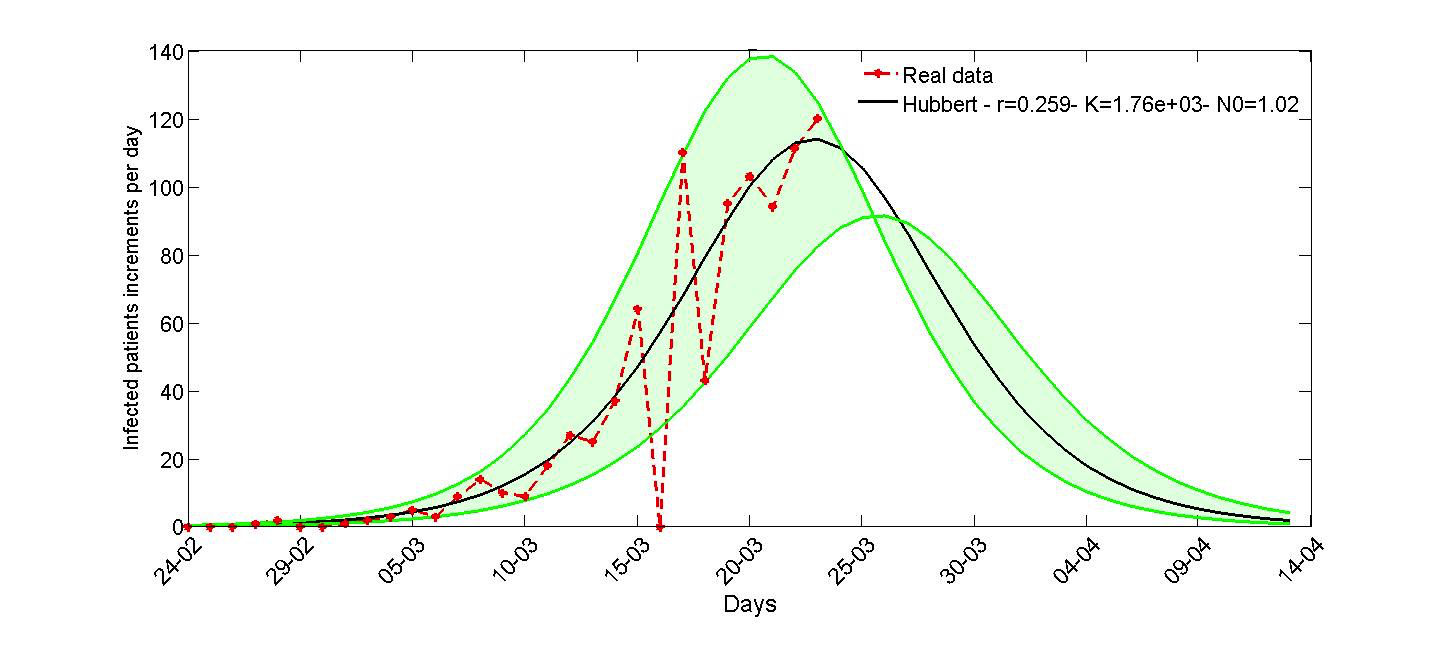}
		\caption{``Regione Puglia".}
		\label{fig:codfig11n}
	\end{subfigure}
\begin{subfigure}{0.49\textwidth}
		\centering
		\includegraphics[width=.9\linewidth]{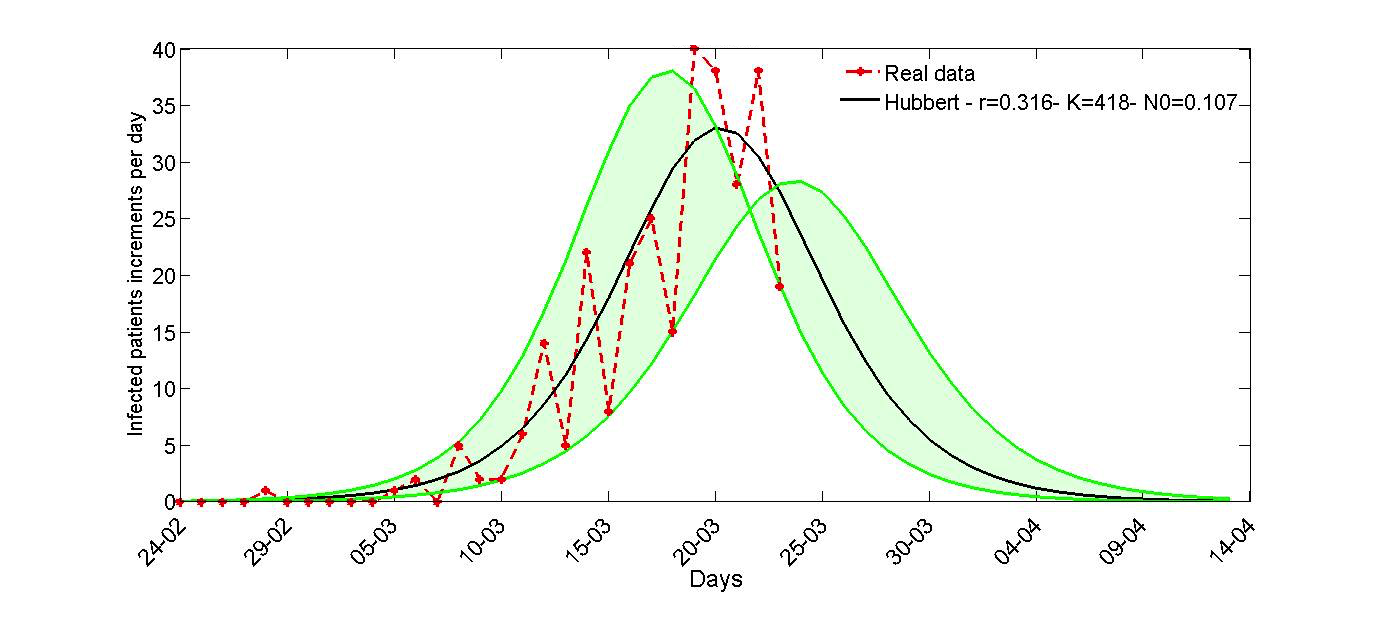}
		\caption{``Regione Calabria".}
		\label{fig:codfig11o}
	\end{subfigure}
\begin{subfigure}{0.49\textwidth}
		\centering
		\includegraphics[width=.9\linewidth]{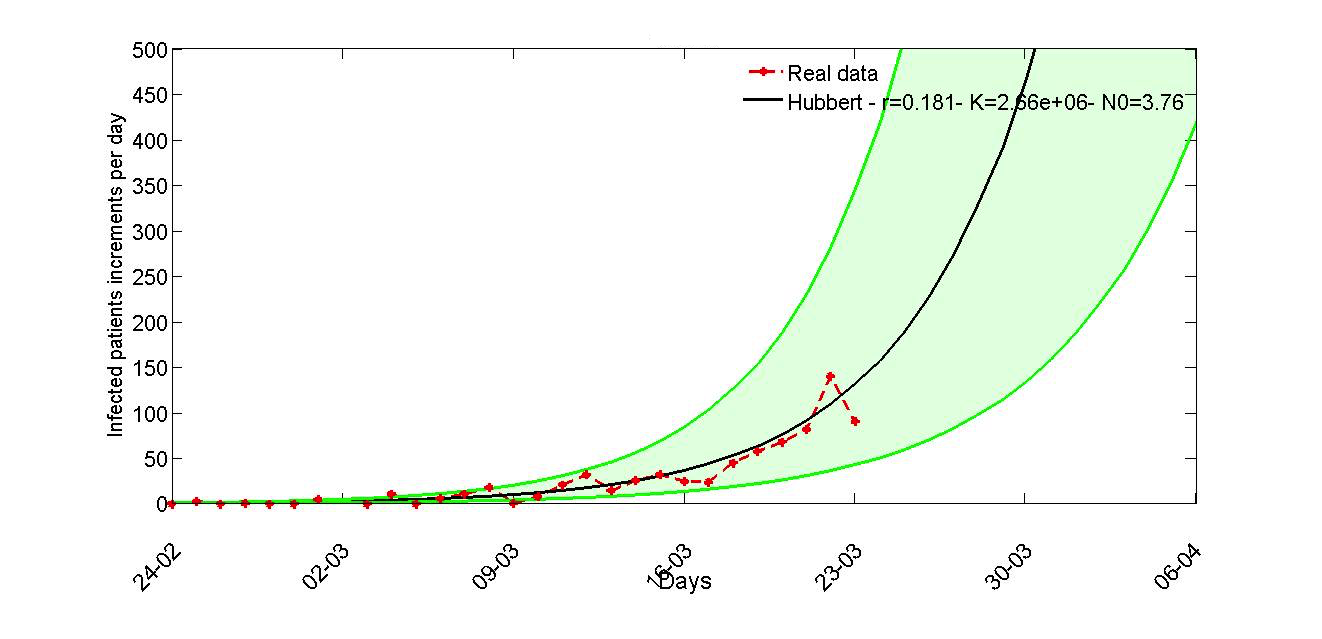}
		\caption{``Regione Sicilia".}
		\label{fig:codfig11p}
	\end{subfigure}
	\caption{Selected sample of regions data and Hubbert function with region of $95\%$ of confidence.}
	\label{fig:codfig11ter}
\end{figure}

\noindent In Figs. \ref{fig:codfig11}-\ref{fig:codfig11bis}-\ref{fig:codfig11ter} we have reported the Hubbert curves of a selected sample of regions\footnote{Data from ``Protezione Civile" https://github.com/pcm-dpc/COVID-19/blob/master/dati-regioni/dpc-covid19-ita-regioni.csv .}.
The plots display an almost coherent scenario with a slow decrease of the emergency in the next months (May). It is worth noting that regarding some regions (for example Basilicata and Sicilia) the situation is still evolving. The available data do not allow a reliable analysis in terms of Logistic and Hubbert curves (the $95\%$ confidence interval is extremely wide) and no peak emergency can be foreseen.
This forecast may be even optimistic and new outbreak of infections, which may spontaneously germinate if restrictions are not properly followed or if not surveyed cases will emerge as acute diseases.\\

\newpage

\newpage

\begin{figure}[h]
	\centering
	\includegraphics[width=0.8\linewidth]{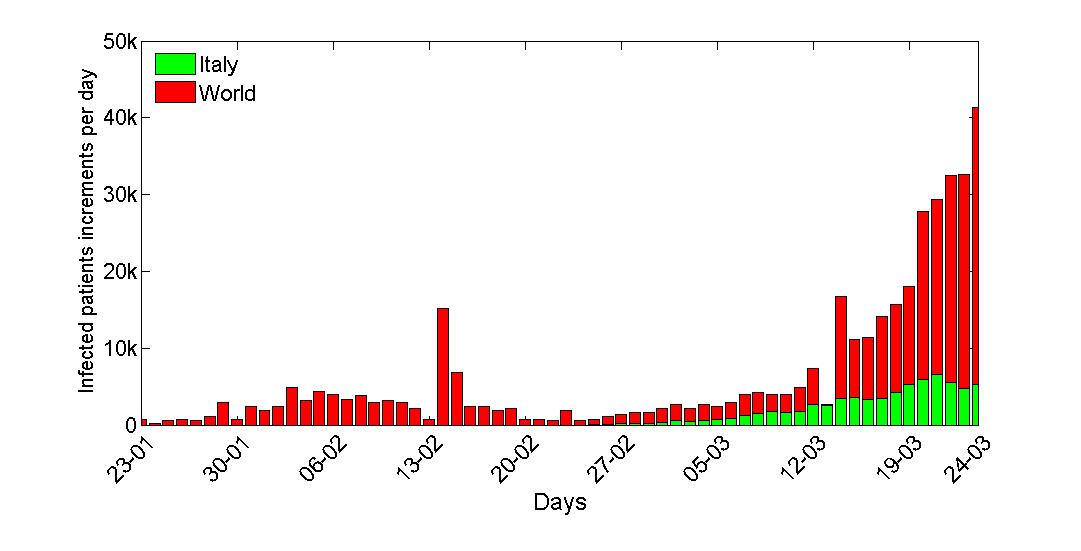}
	\caption{Number of infected individuals since January 23, 2020, in the World and in Italy.}
	\label{fig:codfig12}
\end{figure}

A final element of discussion comes from Fig. \ref{fig:codfig12} where we have reported the worldwide and Italian evolution of the Covid cases/day\footnote{Data from https://coronavirus.jhu.edu/map.html}, a kind of bi-logistic pattern is evident, which supports the ideas put forward in this and in the previous note.\\

What we have attempted here is a little more than the picture of the situation, the lesson we may learn from the present pandemia is important but will be completely understood when not only Italian but the worldwide pattern will be clarified. Probably long time after the end of emergency.\\

\newpage

\textbf{Acknowledgements}\\

The work of Dr. S. Licciardi was supported by an Enea Research Center individual fellowship.\\
\indent The Authors express their sincere appreciation to Dr. Ada A. Dattoli for her help in understanding the biological basis of the infection.\\


\end{document}